\documentclass{vgtc}                 

\ifpdf
  \pdfoutput=1\relax                   
  \usepackage{graphicx}                
  \DeclareGraphicsExtensions{.pdf,.png,.jpg,.jpeg} 
\else
  \usepackage{graphicx}                
  \DeclareGraphicsExtensions{.eps}     
\fi%

\graphicspath{{figures/}{pictures/}{images/}{./}} 

\usepackage{microtype}                 
\PassOptionsToPackage{warn}{textcomp}  
\usepackage{textcomp}                  
\usepackage{mathptmx}                  
\usepackage{times}                     
\usepackage{cite}                      
\usepackage{tabu}                      
\usepackage{booktabs}                  
\usepackage{enumitem}
\usepackage{cite}

\onlineid{1811}

\vgtccategory{Research}

\vgtcinsertpkg

\usepackage{xcolor}
\definecolor{mygreen}{RGB}{34,139,34}
\newcommand{\Add}[1]{\textcolor{black}{#1}}

\newcommand{\Change}[1]{\textcolor{black}{#1}}
\title{ConnectVR: A Trigger-Action Interface for Creating Agent-based Interactive VR Stories}

\author{Mengyu Chen\thanks{e-mail: mengyuchen@ucsb.edu}\\ %
\and Marko Peljhan\thanks{e-mail:peljhan@mat.ucsb.edu}\\ %
\and Misha Sra\thanks{e-mail:sra@cs.ucsb.edu}}
\affiliation{\scriptsize University of California Santa Barbara}

\teaser{
 \includegraphics[trim={0 0 0 3cm}, clip, width=1\textwidth]
 {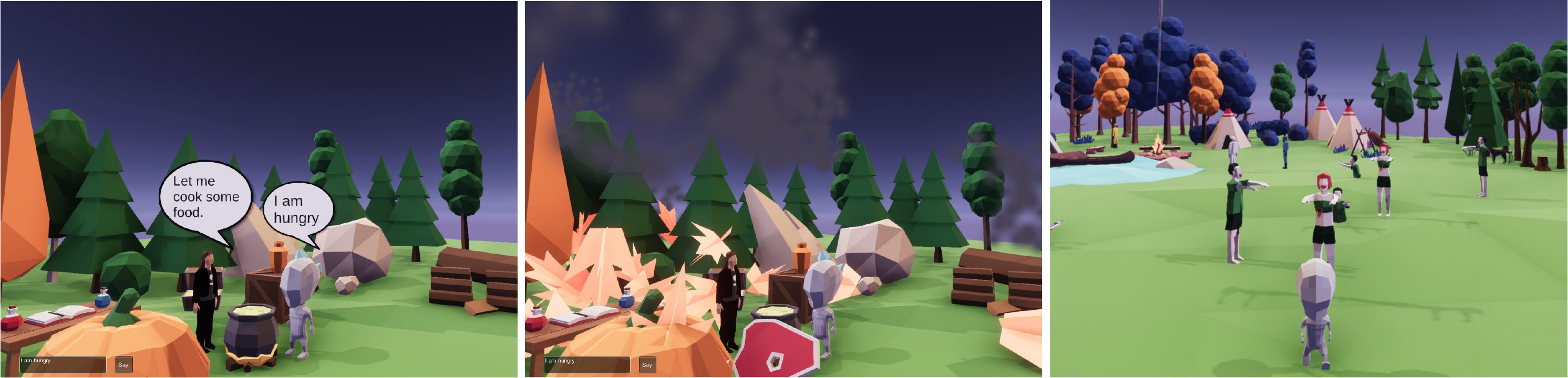}
 \centering
 \abovecaptionskip=1pt
 \caption{Our no-code trigger-action interface focuses on authoring narratives driven by a series of cause-effect relationships triggered by the player's actions. The example here shows story events initiated by the player. Left: (a) the player (low poly avatar) used voice-input to say ``I am hungry'' in an attempt to talk with the virtual character in black, resulting in the non-player character (NPC) responding and starting to cook food. Middle: (b) continued from (a), the NPC's cooking behavior caused a huge fire, and burned down the area. Right: (c) the burn leads to zombies emerging which the player accidentally disturbs prompting a chase.}
 \label{fig:teaser}
}

\abstract{
The demand for interactive narratives is growing with increasing popularity of VR and video gaming. This presents an opportunity to create interactive storytelling experiences that allow players to engage with a narrative from a first person perspective, both, immersively in VR and in 3D on a computer. However, for artists and storytellers without programming experience, authoring such experiences is a particularly complex task as it involves coding a series of story events (character animation, movements, time control, dialogues, etc.) to be connected and triggered by a variety of player behaviors. In this work, we present ConnectVR, a trigger-action interface to enable non-technical creators design agent-based narrative experiences. Our no-code authoring method specifically focuses on the design of narratives driven by a series of cause-effect relationships triggered by the player's actions. We asked 15 participants to use ConnectVR in a preliminary workshop study as well as two artists to extensively use our system to create VR narrative projects in a three-week in-depth study. \Change{Our findings shed light on the creative opportunities facilitated by ConnectVR's trigger-action approach, particularly its capability to establish chained behavioral effects between virtual characters and objects. The results of both studies underscore the positive feedback from participants regarding our system's capacity to not only support creativity but also to simplify the creation of interactive narrative experiences.} \Add{Results indicate compatibility with non-technical narrative creator's workflows, showcasing its potential to enhance the overall creative process in the realm of VR narrative design.}
} 

\CCScatlist{ 
  \CCScat{H.5.1}{INFORMATION INTERFACES AND
PRESENTATION (e.g., HCI)}{Multimedia Information Systems}{Artificial, augmented, and virtual realities};
}

\begin{document}


\firstsection{Introduction}
\maketitle

Interactive narrative design has been widely explored since the emergence of interactive fiction (IF) in the 1970s and hypertext fiction in the 1990s. With simple input methods such as text or mouse clicks, players could explore and interact with the fictional worlds of games like Zork \cite{lebling1979zork} or Varicella \cite{varicella} to control story characters, engage with the setting and make decisions that shape the narrative.
More recently, demand for interactive narratives with expressive visuals has seen a growth with increasing popularity of video gaming \cite{wei2010, cardonarivera2020} and virtual reality (VR). This increased demand coupled with recently introduced technologies in artificial intelligence (AI) presents an opportunity to create new types of interactive fiction that allows players to engage not only with the virtual world but also with AI-empowered virtual characters and narratives from a first person perspective in VR.

In many traditional IF stories, such as Galatea\cite{davila2000galatea}, Façade \cite{mateas2003faccade} or A Dark Room \cite{darkroom}, the player controls the protagonist using a combination of actions commands such as \textit{look, give, approach, build} or natural language via text input. For example, in A Dark Room, the singular action of \textit{stoke fire} has various direct consequences depending on where in the narrative that action is taken (e.g., wood store depletes, a wanderer knocks at the door, some animals approach). The action leads to subsequent newer related actions becoming available to the player (e.g., investigate, turn wanderer away, build a trap). The combination of actions (player clicks on them to trigger the action) and consequences communicated to the player via text descriptions, allows the narrative to unfold uniquely for each player.

These part-game part-story systems are capable of handling player input, assembling the narrative from an array of pre-scripted story fragments, and making everything fit together to create a coherent narrative experience that has the potential to be unique for each player. In contrast, an equivalent narrative in VR needs more detail in the setting and story fragments including sounds and animations along with interactions between the player and the characters. 
Thus, a critical aspect of authoring IF for VR is the need to build explicit causal connections between events and objects along with corresponding visual and spatial representations of outcomes as opposed to the text descriptions used in traditional IF. Existing development methods for creating interactive narratives in 3D and VR are largely adhoc. They either follow game development practices which are not well aligned with the purposes of narratology \cite{cardonarivera2020, silva2019}, or only focus on text-based dialogues for characters without providing support for designing the entire experience \cite{ink, storyassembler, expressionist}. Building cause-effect relationships between player actions and virtual entity behaviors requires substantial programming experience and expert domain knowledge such as familiarity with plot graph construction using logic operators or implementation of AI planning algorithms. Creators such as artists, writers, and storytellers, who may not have the requisite technical skills, are often at a disadvantage with limited options available to them to create interactive VR narratives.

In this work, we present ConnectVR, a trigger-action authoring interface to enable non-technical creators to design VR narrative experiences. Our no-code authoring method specifically focuses on the \textbf{design of narratives driven by a series of cause-effect relationships} (e.g., butterfly effect \cite{lorenz1972predictability}) triggered by the player's actions. In addition to creating the narrative, our interface allows realtime visualizations to ease debugging and support fine tuning. 
Imagine the following scenario where a player is camping in a wooded area and can perform activities such as lighting a campfire, cooking food or firing a gunshot. Reasonable explicit outcomes of these actions could be nearby animals fleeing at the sound of the gunshot or the same animals approaching the camp enticed by the smell of food. A less explicit possible outcome could be the player accidentally starting a forest fire that forces the animals to run and escape. As seen above, player activities can cause both direct and indirect effects on the surrounding virtual agents and the environment, leading to different permutations of story outcomes. To build such a scenario, instead of creating a discrete set of choices and hard coding all possible outcomes \cite{Gordon2004BranchingSI}, a creator can easily build the cause-effect connections between the actions and the agents using nodes in our visual interface. Behind our visual authoring interface is a trigger-action graph composed of action definitions that can make spatial and temporal triggers of consequent actions to be performed by virtual agents, thus establishing various possible relationships and narrative outcomes between the player's actions and virtual agent behaviors. Figure \ref{fig:teaser} shows an example story scenario enabled by our authoring system. 

The main contributions of our work are as follows:
\begin{itemize}[leftmargin=*, noitemsep, topsep=0pt]
    \item A trigger-action based method that centers around causality and spatial and temporal patterns as  authoring guides for non-technical creators of interactive VR narratives.
    \item A no-code visual interface and visualization system to support fast and easy composition of 3D and interactive agent behaviors with complex inter-object relationships.
    \item Results from a workshop study with 15 participants and a three-week in-depth study with two artists \Add{demonstrating how our trigger-action approach opens up new creative opportunities and how it is compatible with the existing artist workflows}.
\end{itemize}

\section{Related Work}
ConnectVR is inspired by prior work in development tools for IF and narrative games,  computational narrative methods \Add{and AR/VR authoring tools for interactive behaviors}.

\subsection{Development Tools for Interactive Fiction and Narrative Games}
There are a number of development systems or programming languages dedicated to creating IF such as Inform 7\cite{inform7}, ink \cite{ink}, and Twine \cite{twine}. Inform 7 is a programming language and an authoring environment targeted at writers and specifically designed for creating IF. It allows writers to use natural language statements to create text and image based stories with different levels of complexity. Ink is a markup scripting language for non-technical writers to create choice-based interactive stories. Other than these tools, automatic story generation using large language model (LLM) is another active area of research. These methods focus on natural conversation generation and computational creativity where artificial intelligence (AI) \cite{park2023generative, ammanabrolu2020} and psychological models \cite{chaturvedi-etal-2017-story, Prez2001MEXICAAC} mimicking human behavior are often explored \cite{alhussain2021}. However, most of these fully automatic approaches are not yet integrated into the development process of actual interactive narratives. 

Video game stories and VR stories, on the other hand, have more requirements for specific hardware device support and realtime graphics. These stories are often created using a game development pipeline created using engines such as Unity \cite{unity} or Unreal \cite{unreal}. However, game engine workflows are often complex for non-technical creators and require software engineering skills and programming experience. 
A number of systems have been specifically proposed to aid the development of interactive game narratives such as Expressionist \cite{expressionist}, StoryAssembler \cite{storyassembler}, IFDBMaker \cite{ifdbmaker}, StoryPlaces \cite{storyplaces}, Ensemble Engine \cite{Samuel2015TheSR}, Comme il Faut \cite{McCoy2011CommeIF}, and Villanelle \cite{villanelle}. However, similar to the tools available for IF creators, these tools also do not provide an integrated solution or full pipeline that non-technical creators can easily use to build their own narratives. For example, Expressionist is a web-based in-game text generation tool that can help generate descriptive texts for virtual objects based on attribute tags and character dialogues and conversations. The Ensemble Engine is a visual authoring tool that can help define and simulate social factors that impact virtual character relationships and their motivations using scored rules. Villanelle proposed a blockly-based \cite{blockly} programming language \cite{blockly} to simulate autonomous character behaviors.

ConnectVR simplifies the complex game development workflow and its no-code interface does away with the need for scripting familiarity in prior works. Non-technical creators can create interactive VR narratives using a visual interface for composing complex relationships and behaviors. To support meaningful interactions in a story, in addition to what previous tools enable, our system allows the creation of avatar actions driven by a variety of player inputs, including voice input, VR controller events, as well as natural body gestures in VR such as position and movement in space, gaze and touch. The wide variety of input methods allows for flexibility in developing for immersive environments.

\subsection{Computational Narrative Methods}
Building a computational narrative model that facilitates meaningful selection of narrative events based on player interaction is a challenge \cite{young2013review}. Many prior works have focused on plan-based narrative generation methods and have proposed various interactive systems that take factors such as character personality \cite{orkin2003applying, Bahamon_Barot_Young_2015}, dilemma \cite{gardin2009}, social relationships \cite{porteous2013}, intent \cite{riedl2006}, task hierarchy \cite{htnplanner,cavazza2002}, and authorial goals \cite{Riedl2009IncorporatingAI}, to model logical and coherent narratives \cite{riedlandyoung2010, young2013review}. For example, Goal Oriented Action Planning (GOAP) is a decision making architecture widely used in video games to help NPC characters plan and adjust their behaviors to reach preset goals defined by the author \cite{orkin2003applying, orkin2006}. Interactive Behavior Tree (IBT) extends upon the Behavior Tree \cite{halo22005} formalism's advantage in authoring branching narratives to handle free-form player interaction in the story \cite{kapadia2015, kapadia2015evaluating}. 
However, the complexity in the specification of these systems can hardly be translated into an easy authoring process for non-expert creators, as it requires specific domain knowledge such as writing Planning Domain Definition Language \cite{fox2003pddl2}) or plotting behavior trees with logical operators \cite{riedl2012openproblems}. With a goal to flatten the learning curve while maintaining a high level of expressiveness, our proposed method uses a "trigger-action" user conceptual model \cite{Dey2006} often used in IoT context-aware policy programming to guide cause-effect interactions between the end-user and smart objects under different contexts. Creators do not need prior domain knowledge or understanding of logic operators to build the cause-effect chains. Our simulation-based approach \cite{riedlandyoung2010} involves autonomous agents that are capable of performing actions triggered by player behaviors once the creator defines them using the visual interface. 

\subsection{\Add{AR/VR Authoring Tools for Interactive Behaviors}}
\Add{There is an increasing demand for AR/VR authoring tools that can provide dynamic and interactive interfaces to ease the creation of interactive and immersive behaviors \cite{Ashtari2020}. One prominent approach is Programming by Demonstration (PbD) \cite{myer1986,myer2000}, where users demonstrate movements or interactions to a system that can generate code or actions for desired end-user behaviors. The demonstration process often involves a series of acting and capturing to ensure users can clearly communicate their intent while generating AR or VR prototypes \cite{leiva2020, Leiva2021, sayara2023, Arora2019}. For example, RAPIDO provides a no-code prototyping system that allows designers to use a mobile AR device to record a video prototype to capture touch inputs, animation paths and rules, ultimately producing an interactive AR prototype based on a state machine generated by PbD \cite{Leiva2021}. Sayara et al.\cite{sayara2023} proposed an immersive prototyping system that uses PbD along with event-driven state machines and trigger-action authoring to enable users to design, test and deploy Compound Freehand Interactions in VR. Direct manipulation, often used to create screen-based interactive animations \cite{li2005, ledo2019, nicholas2023, kazi2014, kazi2014kitty}, offers an alternative to PbD to author AR/VR interactions. Recent systems have enabled the use of hand-gestures \cite{eitsuka2013, Arora2019}, sketching \cite{wacker2019, gasques2019} and physical objects \cite{nebling2019} to create and control different virtual objects. While both PbD and direct manipulation are highly useful in authoring user-driven interactions, the ceiling of expressivity of the created experience is often limited by the input modality of the system \cite{Leiva2021}. Conversely, Visual Programming (VP) systems allow users to create logical behaviors and interactive scenes via graphical representation of data-flow and function operation without extensive scripting, offering a higher expressive ceiling \cite{zhang2020, wang2021, ens2017, chen2021, chen2023}. Flowmatic is a reactive VP system that allows users to build fully interactive experiences with immersive authoring \cite{zhang2020}. GestureAR is an authoring tool for creating interactive AR experiences, featuring VP and embodied demonstration capabilities for custom design of freehand interactions \cite{wang2021}. Game engines like Unity and Unreal also provide their own VP interfaces to support visual composition. However, unlike no-code methods, most VP systems require the users to have sufficient knowledge of programming concepts (e.g., variables, macros, state-machines). Prior works have shown that, artists lacking familiarity with programming often avoid reading or manipulating numerical data in symbolic programs, and can feel overwhelmed by the amount of numerical information displayed on screen. Consequently, working with visual programs driven by numerical data or algorithmic symbols remains a challenge for such creators \cite{li2020, macullough1996}. 
}

\Add{To overcome some of the limitations for artists, ConnectVR adopts a natural language style trigger-action visual graphing method, enabling creators to compose interactive and cause-effect behaviors in a virtual environment. It is tailored for crafting VR narratives with complex and abstract inter-object relationships which PbD and direct manipulation systems do not enable. Our trigger-action based visual graph is defined solely by actions and connections, making it much simpler to learn than existing VP systems. This simplicity is achieved by removing the need to have prior knowledge of fundamental programming concepts like variables, state-machines, or logic operators.}

\begin{figure*}
  \includegraphics[width=0.80\textwidth]{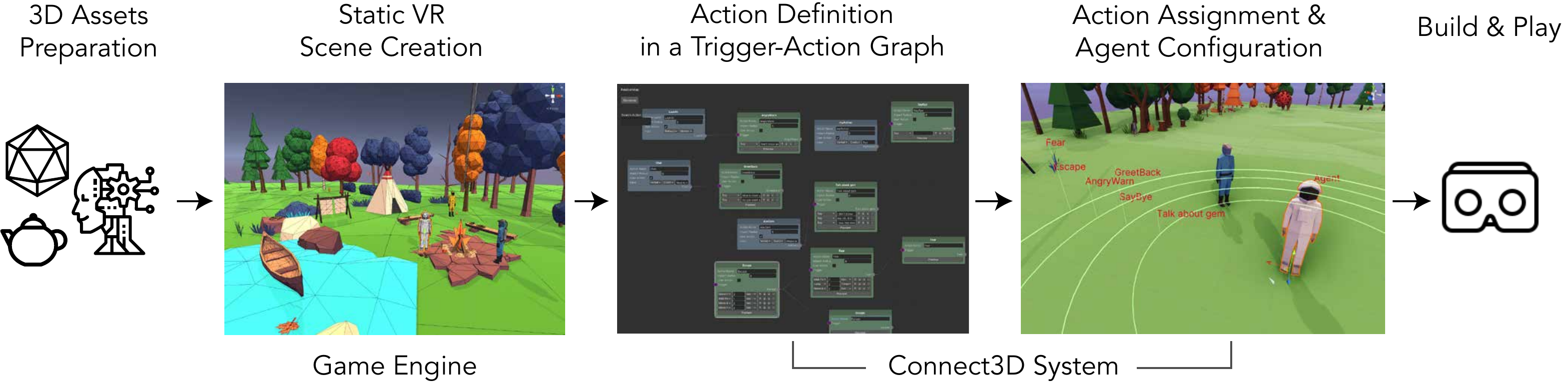}
  \centering
  \abovecaptionskip=0pt
  \belowcaptionskip=-12pt
  \caption{Overview of the creator workflow using our ConnectVR authoring interface in the Unity Game Engine. After setting up a static 3D scene, a relational graph composed of actions and relationships allows creators to design and assign actions to the virtual agents in the 3D scene. }
    \label{fig:workflow}
\end{figure*}

\section{System Design}
We built ConnectVR as Unity game engine plugin with a custom visual authoring interface to support fast and easy no-code creation of 3D narratives. Our system specifically focuses on creating narratives driven by a set of creator-defined trigger-action rules for virtual agents to render complex cause-effect relationships between a player's actions and corresponding story events. \Change{In this section, we outline a set of design goals, which guided our system development, and introduce an authoring workflow of our "trigger-action" approach for linking player input in VR with smart agent behaviors to create meaningful and personalized VR narratives.}

\subsection{Terminology}
In this work we use the following definitions of these terms:
\begin{itemize}[leftmargin=*, noitemsep, topsep=3pt]
    \item \textbf{User / Creator} - the artist or storyteller who desires to build interactive narratives in 3D (on screen) or VR using ConnectVR.
    
    \item \textbf{Player} - the human who goes through the created narrative experience in 3D or VR.
    \item \textbf{Agent} - the virtual entities in the story that are capable of interactive and dynamic behaviors, including but not limited to characters, scene objects, particle effects, etc.

    \item \textbf{Static Object} - the story background objects that do not change or have dynamic behaviors, the opposite of agent. This includes the terrain, mountains, and other landscape elements.

    \item \textbf{Action/behavior} - the basic narrative element in the ConnectVR system that represents a behavior-based plot fragment. An action can be performed either by a player or by an agent. Player actions in VR are mapped to their physical input events such as VR controller button press, hand gestures, speech input, etc.

    \item \textbf{Trigger} - The pre-condition of an action. Actions performed by a player are all considered triggers (unless they do not have any impact on the story).
\end{itemize}

\subsection{Design Goals}
Our main goal is to enable non-technical creators (users of ConnectVR system) to design narrative experiences where the consequences of a VR player's actions are persistent and impact how the story progresses. 
Inspired by Janet Murray's work on the future of digital storytelling and the idea of \textit{cyberdrama}\cite{janetmurray2004, murray1997}, our goal is to allow users to build a VR story world that is populated by a player (the person experiencing the narrative), interactive virtual agents (e.g., characters, mobile scene objects like cars and animals) and static scene entities (e.g., trees, buildings, terrains). In this world, the player and the agents are capable of performing actions that are the mimetic activities \cite{aristotlepoetics, Louchart2004NarrativeTA} which have consequences to formulate cause-effect linked narrative plots. The creator's task is to design the actions, and determine each action's causal, temporal and spatial relationships to the player and to the virtual agents. \Change{Building upon Murray's work} \Add{and considering the pros and cons of prior works}, we define the following design goals:
\begin{itemize}[leftmargin=*, noitemsep, topsep=3pt]
    \item \textbf{\textit{Conciseness}}: The goal is to make it easy to compose VR narratives, without the need for any prior coding knowledge. For this, the authoring interface needs to be easy to learn, so that users can quickly create VR narratives with complex agent behaviors that respond to player actions.
    \item \textbf{\textit{Expressiveness}}: The tool should be flexible enough to allow creators to build a variety of branching VR narratives with a wide range of complex plots that can provide personalized experiences to players. 
    \item \textbf{\textit{Intelligence}}: As VR offers unique opportunities for a player to naturally interact with the virtual environment, the system should be able to support various kinds of natural player input methods, such as voice and gestures. For voice-based interaction, it would be highly desirable to integrate recent AI models (e.g., ChatGPT) to enable more natural conversations between the player and the agents.
    \item \textbf{\textit{Connectivity}}: The system should integrate with the existing ecosystem of tools to streamline the creator's workflow.
\end{itemize}

\subsection{Authoring Workflow}
While our system can output both on-screen 3D narratives and VR narratives, in the paper we focus on VR narrative design. The player experiences this by putting on a VR headset (HMD) and interacting with the virtual world through handheld controllers, voice input, body movement (e.g., head turn, walking) and gestures. \Change{Our system provides a visual interface for non-programmers to create sequences of trigger-action events, which make it easy to compose complex, cause-and-effect narratives. While it's possible to create similar sequences with Unity's visual scripting, our authoring method introduces a higher level of abstraction from low-level functionalities that is closer to natural language descriptions.} \Add{Instead of constructing the building blocks based on numerical data-flow which can be overwhelming for non-technical users \cite{li2020, macullough1996}, we adopt a conceptual model of \textit{action-flow} to allow creators to seamlessly build intricate cause-effect behaviors. This extended trigger-action pattern is designed to facilitate relationship building between player and agent behaviors, steering the creator's focus away from technical implementation towards narrative behavior-centered design thinking.}


\autoref{fig:workflow} shows an authoring workflow for how a user can compose a VR narrative. Users begin by importing and placing 3D models and assets in the Unity scene editor to compose the static story environment. 
After composing a desired static scene, the user opens up the ConnectVR authoring interface to define an \textit{action graph} that is made of individual action blocks (e.g., player input behaviors, virtual agent behaviors) and their trigger-action patterns (e.g., action A triggers action B, and action B triggers action C and so on). The user can preview the consequences of their assigned actions in the simulation preview mode to help them fine tune the outcomes. 
\begin{figure}
  \includegraphics[width=0.48\textwidth]{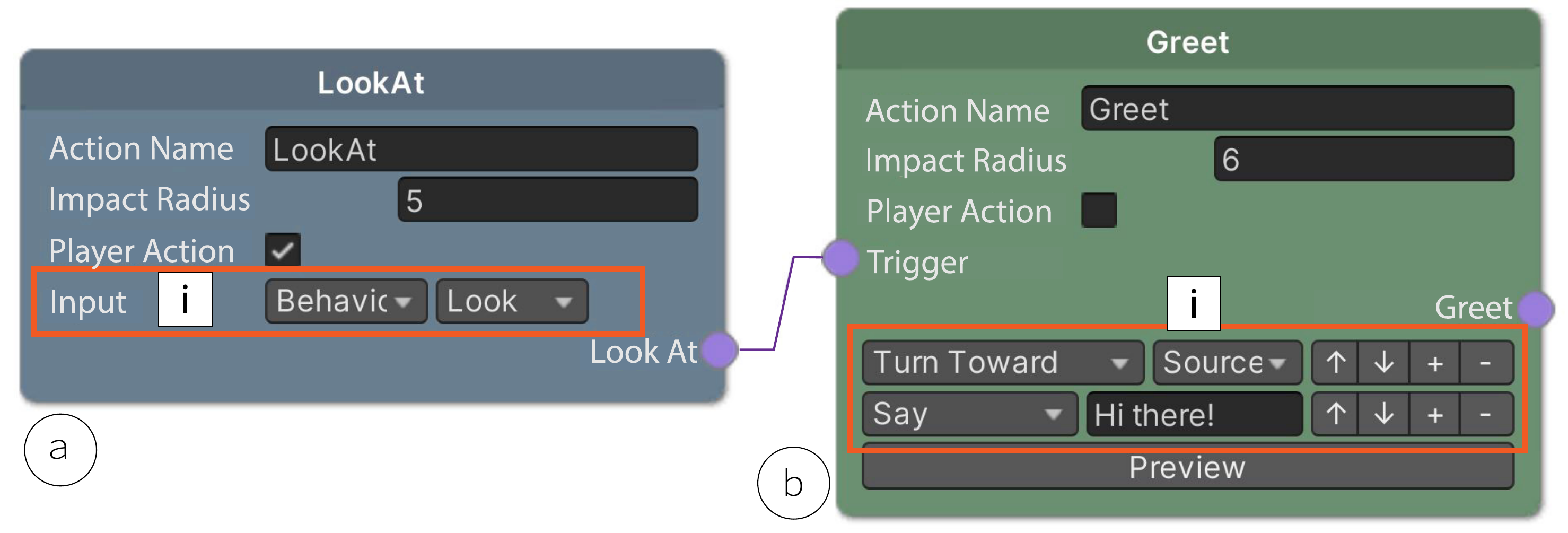}
  \centering
  \abovecaptionskip=1pt
  \belowcaptionskip=-12pt
  \caption{An example of how actions are represented and linked in the ConnectVR visual interface. Left: (a) a player action is represented as a blue colored node, in which the creator can select a behavior mapping to a player input event (e.g., player's `Look' gesture). Right: (b) an agent action node, represented in green, include both input (left edge of the node) and output ports (right edge of the node) that take causal links. The example node here has (i) an execution block where the creators define animations or state changes to be played and rendered on the virtual agent. The linkage between the nodes ensures that the agent action on the right triggers whenever the player looks at the agent from within a 5 meter range.}
    \label{fig:actionnode}
\end{figure}
\begin{figure*}[t]
  \includegraphics[width=0.85\textwidth]
  {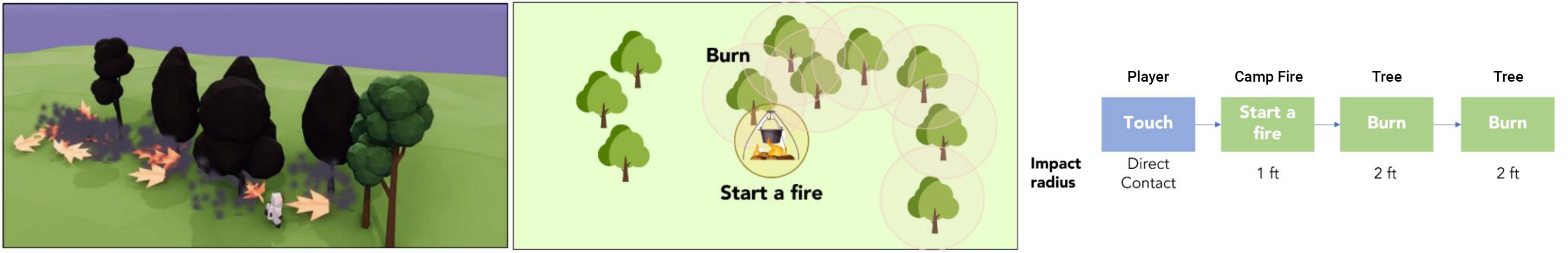}
  \centering
  \abovecaptionskip=1pt
  \belowcaptionskip=-12pt
  \caption{An example of how spatially constrained cause-effect relationships may affect the narratives. Left: a scenario where the player accidentally burned down the entire wood area. Middle: a spatial map showing the location of different agents and the impact radius of each action. A sequence of “burn” becomes possible when the trees are arranged closely after the initial “start a fire” by the campfire. Right: a trigger-action sequence leading to a fire in the wood. The player initiates a “touch” action on a campfire to “start a fire” on the campfire. Once “start a fire” is activated, any nearby tree object capable of “burn” action within the campfire’s 1 ft impact radius will trigger the “burn” action. Subsequently, this “burn” action may trigger other trees' “burn” action within this tree’s 2ft impact radius as defined for “burn” action.}
    \label{fig:spatialmap}
\end{figure*}
\subsection{Block-based Action Model}
We adopt an action-based model to represent the narrative plot --- both the player and the agents are capable of performing actions and each action has consequences which can trigger further actions. The player's experience is composed of sets of actions initialized and executed based on the choices they make as they play through the narrative. An action, in other words, is the basic narrative element that represents a behavior-based plot fragment. The creator defines each action as an abstraction for what a player, embodied in an avatar, or a non-player agent does. For example, an agent can perform a \textit{Cook} action which involves a series of animated events such as moving to the kettle and starting a fire. Similarly, a player action is mapped to their voice, gesture, or controller input events. The creator can define what type of player input would be considered valid actions in the VR story world by specifying input parameters (e.g., controller button key, speech content or topic, specific natural gesture). For example, the creator can define a \textit{LookAt} action by mapping the VR player's head turn to looking at a virtual object or agent. During runtime, whenever the player looks at the pre-defined target object, this \textit{LookAt} action will become active. 

\subsection{Causal Relationship}
When an action is in an active state (e.g., the player is talking, a virtual agent is dancing), this action may serve as a trigger for another action, based on the creator's defined causal relationship. In the trigger-action graph, the creator can draw a causal link between two actions, which relates the effect of a trigger to the pre-condition of a consequent action. \autoref{fig:actionnode} shows an example of how two actions are linked. This method for causality builds upon a simple idea: every action can have multiple consequences which can trigger subsequent actions leading to further consequences and so on. Typically, a narrative plot would start with a player action that is the main trigger of many consequent agent actions. However, it is possible to design fully autonomous agents that have pre-defined actions which are the initial triggers of other consequent actions, thereby beginning a narrative experience without need for immediate player intervention.

\subsection{Spatiotemporal Constraints}
To provide greater control over the causal relationship between a trigger and its consequent actions, we provide extra modulation by inclusion of spatial and temporal factors, which are the position and availability of agents and the impact radius of each trigger. \autoref{fig:spatialmap} shows an example of how spatially constrained cause-effect relationships may be created. When a player or an agent has performed a trigger, the effect is only valid within a pre-defined spherical radius around the origin location. If no other agent is nearby, then this trigger would not lead to any consequences. Increasing the impact radius can ensure the activation of consequent actions but it may also lower the plausibility of the cause-effect relationship in some cases (e.g., a speech made by a virtual agent in a room should not be heard by others outside of the room). There is also a customizable cool-down timer that can be set by the creator to prevent the same trigger from happening again for some period of time. In this way, the creator can more precisely control each action and its fallout.

\subsection{Relational Play}
\begin{figure*}[t]
  \includegraphics[width=0.85\textwidth]
  {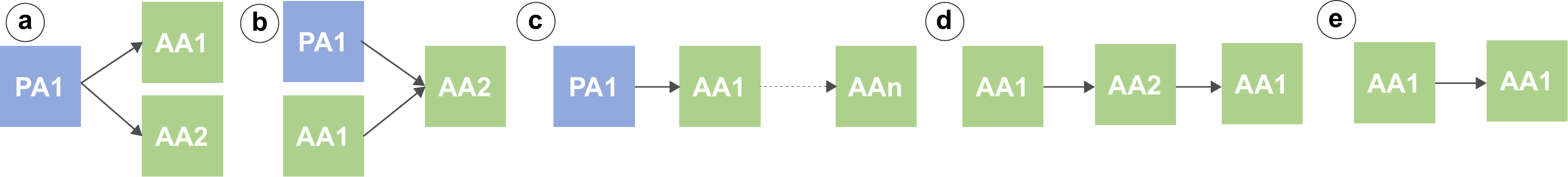}
  \centering
  \abovecaptionskip=1pt
  \belowcaptionskip=-13pt
  \caption{Examples of possible relational play by the creator. (a) one player action may lead to multiple other agent actions. (b) multiple different actions may lead to the same result agent action. (c) an action chain can be made of any length to formulate chained reactions. (d) an agent action can be the cause of its causal action to formulate a feedback loop. (e) an action can be the causal action of itself to formulate a spreading effect.
}
    \label{fig:relationalplay}
\end{figure*}

As the development of a story can be highly creative and may involve various types of cause-effect relationships \cite{raskin1998}, we enable the composition of a flexible trigger-action graph so that there are very few restrictions on how a creator may design and build the causal chains. Just like real life, in the graph, a trigger may lead to many subsequent actions to simulate chained reactions or butterfly effects.
A feedback loop can occur where a trigger causes an action which results in the trigger causing the same action again
creating an infinite entanglement of multiple agents over time and space. \autoref{fig:relationalplay} shows a set of possible ways to design trigger-action patterns.

\subsection{Player Action Input Methods}
To support natural interactions, ConnectVR provides various player input methods that creators can map to create player actions. Our system categorizes these input methods into three categories:
\begin{itemize}[leftmargin=*, noitemsep, topsep=3pt]
    \item \textbf{\textit{Device Input}}: various hardware device input can be mapped to player action such as mouse click, keyboard key-press, or VR controller button-press.
    \item \textbf{\textit{Natural language Input}}: we support microphone-based voice input for VR player to engage in a natural conversation with virtual agents. The creator can choose to use our built-in intent parsing feature (based on UnityNLP library) to summarize the sentence input into a specific category (e.g., greetings, wayfinding, identification) or use the sentence as a prompt to trigger AI-driven responses from an agent supported by ChatGPT's text generation and Azure Speech Service's voice synthesis. 
    \item \textbf{\textit{Body Gesture input}}: A VR player's head pose, body gestures or locomotion can also be mapped to narrative actions such as looking at someone or something, picking up an object, moving toward a target, touching, waving etc.
\end{itemize}
All these input methods can be defined in the player action node and serve as triggers for agent behaviors, which enable a rich and highly interactive experience. Separating player input methods from player actions in the system enables creators to easily define and map player actions to agent behaviors. This separation allows creators to focus on defining player actions without worrying about the underlying technical details of input event handling.

\subsection{Agent Action Commands and Assignment}
For purposes of re-usability, an agent action is designed as a template that can be assigned to many agents. Since the definition of an agent is not limited to a virtual character but includes any virtual entity (a tree, a car, etc.), an action can also be an abstract representation of an object's state change (e.g., slowing down, gradually deteriorating). 
We categorize the action commands into seven different categories for the ease of organizing and understanding:
\begin{itemize}[leftmargin=*, noitemsep, topsep=3pt]
    \item \textbf{\textit{Spatial Control}}, which includes locomotion and transform related commands such as move, follow, rotate, scale, etc.
    \item \textbf{\textit{Visual Control}}, which includes graphics and rendering related commands such as color change, VFX play and stop, material change, appear or disappear, etc.
    \item \textbf{\textit{Temporal Control}}, which includes time related commands such as standby, delayed or instant trigger (immediately trigger its linked consequent action on nearby agents), freeze/defrost (stop or start taking trigger on this agent), etc.
    \item \textbf{\textit{Language \& Sound Control}}, which includes audio or speech related command such as play sound clip, speak a sentence (the creator can provide a pre-written sentence or feed in a prompt to ChatGPT API for generative text content).
    \item \textbf{\textit{Animation Control}}, which includes animation related commands such as play/stop animation clip, animation state change.
    \item \textbf{\textit{Instance Control}}, which includes generative commands such as spawn object, activate / deactivate agent.
    \item \textbf{\textit{Utility Control}}, which includes general and advanced state control of the agent, and custom function call, etc.
\end{itemize}

All these commands can be mixed and stacked together to form custom agent actions. Creators can build more interactive and complex virtual agents by adding more agent actions. Dragging and dropping agent actions onto 3D objects in the scene composer is a simple way to add interactive capacity to the virtual agents. 
We provide custom interface support to automatically visualize the impact radius of each assigned agent action. This preview helps creators ensure their agent behaviors will be activated as intended. 

\section{Evaluation}
Our evaluation had two main objectives: to evaluate the usability and expressiveness of ConnectVR in an open-ended setting and to understand how our trigger-action method enables non-technical creators to compose VR narratives.
We conducted two studies targeting both objectives. The first was a workshop with 15 participants from creative backgrounds, aimed at evaluating the expressiveness and ease of use. The second was an in-depth three-week study with two creators, aimed at evaluating ConnectVR's performance in a more realistic scenario to reveal any trade-offs between our no-code system and existing tools used by the participants. Building upon the opportunities and challenges in prior works \cite{jacobs2017,jacobs2018, jingyi2021}, we developed the following evaluation criteria:
\begin{itemize}[leftmargin=*, noitemsep, topsep=3pt]
    \item \textbf{\textit{Process}}: Is it easy and comfortable for non-technical creators to use the tool? How does working with the tool compare to existing authoring tools? Can it fit into their current creative process? 
    \item \textbf{\textit{Creative Outcomes}}: Does the tool support the creation of different narrative experience? Were the creators able to achieve what they want to do with the tool?
    \item \textbf{\textit{Reflection}}: Does the tool affect how creators think about creating narrative experience?
\end{itemize}
Collected data included quantitative and qualitative surveys, reflection write-ups, observation videos, recorded interviews, and the narratives created by the participants. 
Surveys for the workshop study contained questions regarding the participants prior experience, utility, creativity support of the tool, and overall experience with ConnectVR's authoring workflow. Surveys for the in-depth study focused more on the qualitative aspects of the creative experience, including notes on daily creative goals and achievements, and longer reflection write-ups. Survey results, interview transcripts, and the projects were analyzed based on our evaluation criteria. Before the studies, we conducted a pilot workshop with three participants to get early feedback on ConnectVR’s user interface and functionality to help us identify and fix technical and user experience issues.

\subsection{Workshop Methodology}
We conducted the workshop with 15 participants (5 female, 9 male, 1 undisclosed, age range 18-44) to understand if our prototype meets our design goals. The study took place in-person over three sessions to accommodate schedules and availability. We designed our evaluation to see if participants were, 1) able to use our visual authoring interface to construct cause-effect behaviors between player and virtual agents, 2) able to create original 3D narratives demonstrating expressiveness of our system, and 3) to gain insights into the usability and ease of authoring with our system. We looked for participants with interest or experience in VR narrative or game design. Participants were from backgrounds including media arts (9), computer science (3), film (2), and communication (1). 
A 5-point Likert scale was used for all ratings (1 = not at all, 5 = A great deal). Participants rated an average of 2.86 regarding their familiarity with creating 3D models or 3D scenes, with Blender being the most common 3D software (6/15 participants). They also rated an average of 3.3 on familiarity with creating interactive 3D or VR experiences, with Unity being the most common development tool (6/15 participants). On average, participants rated their computer programming experience at 3.8, with Python being the most common language (9/15 participants). We paid participants \$30 for a 120-minute workshop.
\vspace*{-3pt}
\subsubsection{Procedure}
We instructed participants to install Unity 2020.3 on their Windows 10 laptops before the workshop. This ensured a uniform game engine setup for our ConnectVR package. Each participant received a unique participant ID, provided informed consent (protocol 7-21-0749), and filled out a pre-study questionnaire. They were given a link to download our ConnectVR Unity package. A researcher provided a 25-minute interactive tutorial on ConnectVR, guiding participants through the basics of Unity and the creation of a simple interactive experience. Participants were encouraged to ask questions and try out all the features. The researcher ensured that each participant was able to reproduce the example experience before proceeding with the study task. After the tutorial, participants were asked to create a short narrative in the next 60 minutes. The story needed to have player-driven cause-effect relationships. We provided a low-poly 3D model library that contained over 2000 prefab objects to let participants build their scenes and narratives without spending precious time and effort creating or finding 3D models. To scaffold creativity, we also provided two virtual environments, a wild western town and a forest park, both of which contained a variety of static 3D models (e.g., virtual characters, animals, houses, vegetation). 
The researcher observed the authoring process, and answered any system related questions. Due to time constraints and COVID social distancing practice, we did not provide actual VR headsets to let the participants test the experience they created but they were encouraged to take home the project and test on their own devices (they could still use the simulator or third-person avatar to try out the experiences they built). After the task session ended, participants were asked to upload their projects to a cloud drive and fill out a post-study survey. The questionnaire asked about the usability of ConnectVR, its capacity to support creativity, and the ease and efficiency in creating VR narratives using the visual interface.

\subsubsection{Workshop Results}
\begin{figure}[!t]
\includegraphics[width=0.42\textwidth]{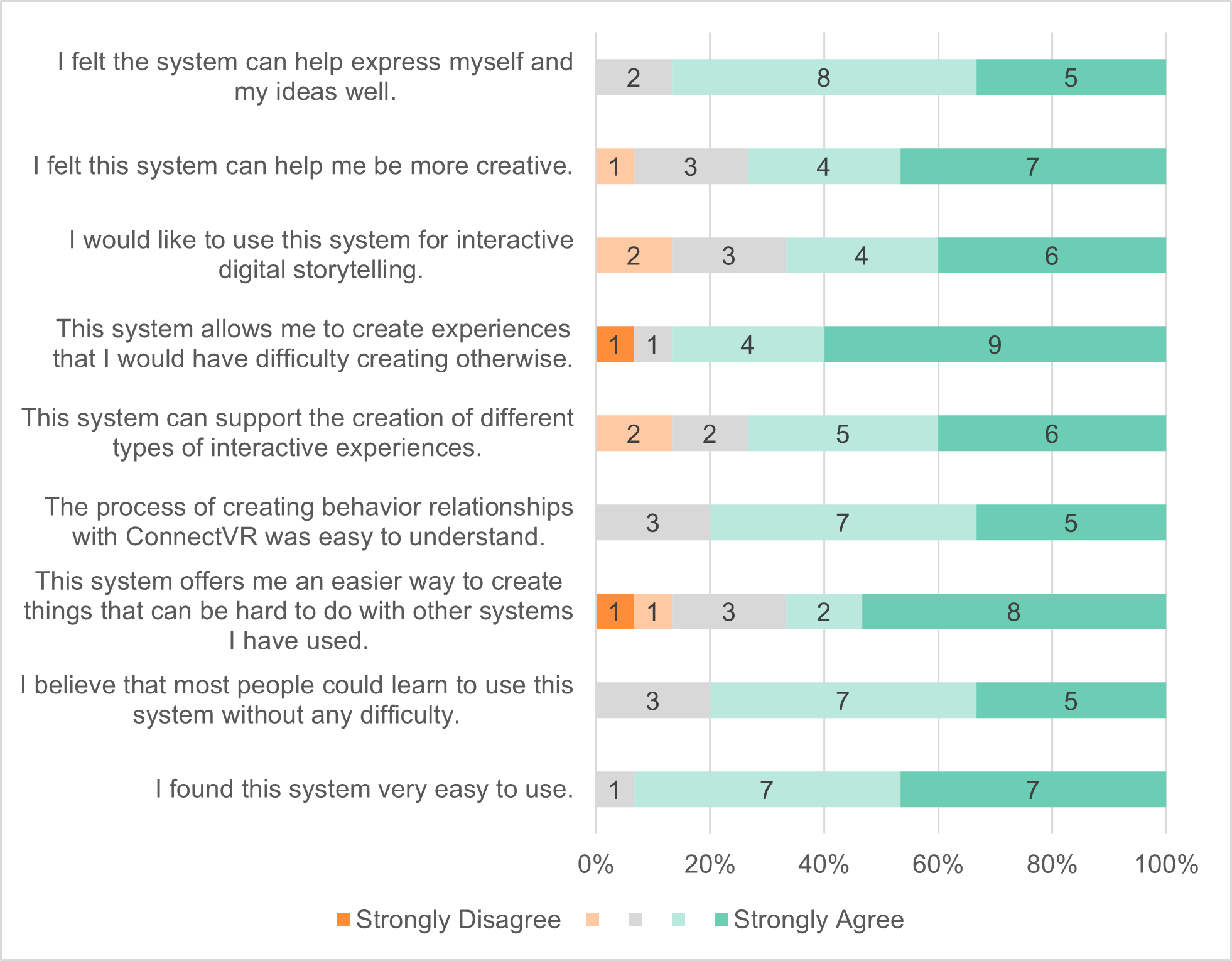}
\abovecaptionskip=1pt
\belowcaptionskip=-12pt
\centering
    \caption{Results from the workshop study survey.}
    \label{fig:usabilityeval}
\end{figure}

\begin{figure*}[t]
  \includegraphics[trim={0 1cm 0 0.2cm}, clip, width=1\textwidth]{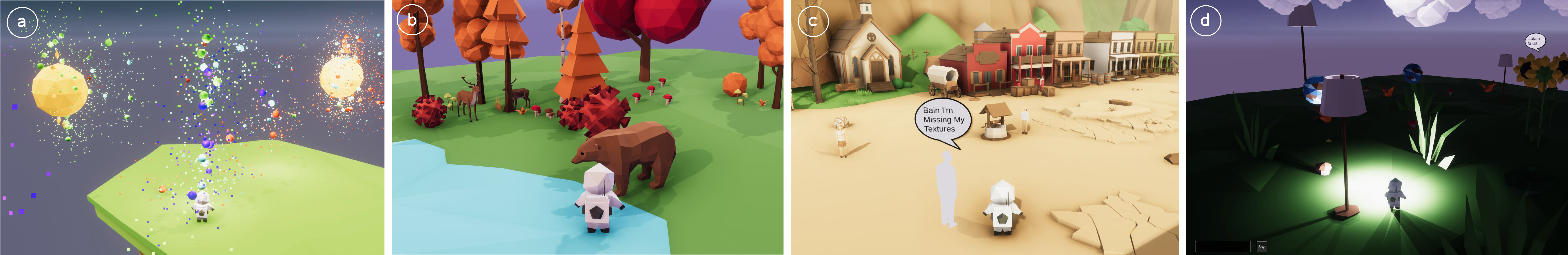}
  \abovecaptionskip=1pt
  \belowcaptionskip=-12pt
  \caption{Example story scenes from the workshop creative task. (a): P1 created an art experience where hundreds of intelligent agents, aware of the player's presence, frequently attach to the player's avatar. Repulsive actions occur between agents of different colors, and they change colors upon collision with large spheres. (b): P9 crafted a forest story where a player encounters a red COVID-like virus. The virus constantly duplicates and spreads when approached by wandering animal or human characters. (c): P13 created a walking simulator experience in the wild western scene, featuring eight different agents with distinct behaviors and dialogues for player exploration. (d): P11 made a fantasy land occupied by unusually oversized flowers and mushrooms where player movement triggers various object behaviors such as dancing, singing, and light flashing.
  }
\label{fig:results}
\end{figure*}

All workshop participants were able to create their own interactive short stories, including those with no prior experience in creating interactive narratives in 3D or in VR (P3, P5, P6, and P7). \autoref{fig:results} shows example scenes from the participant designed narratives. 
Below, we detail the results in the context of our evaluation criteria:

\textbf{Process}: All participants were able to understand the trigger-action concept and applied it to compose cause-effect relationship graphs for virtual agent behavior design. This observation is substantiated by survey responses (\autoref{fig:usabilityeval}): 14 participants reported positive (average score of 4.4) indicating that the system and interface was highly usable and understandable. 12 participants agreed that most people could learn to use the system without any difficulty. 12 participants were positive (average score of 4.1) about the process of creating behavior relationships being easy to understand. 13 participants agreed that the system allowed them to create experiences that they would have difficulty creating otherwise, with one strong disagreement by P11. 

\textbf{Creative Outcomes}: Participants 
designed a total of 96 actions collectively, among which the most popular ones were spatial actions (21.6\%) followed by language and sound-based actions (18.3\%). Overall, participants exhibited interest in designing experiences that were driven by human-agent conversations. For player triggers, participants made 24 different types with the most popular input event being language based input (10). We also observed an interest in making chained cause-effect behaviors that began with player input. In each story, the number of cause-effect chained relationships (those with unconnected head and tail) had a median (\textit{M}) value of 5 and median absolute deviation (\textit{MAD}) of 1. Over 53\% of the relationships involved 3 or more actions and over 38\% had 4 or more than 4 actions chained together. 
In this one-hour task, participants created a total of 147 interact-able virtual agents. The number of interact-able agents in each story had \textit{M = 8} and \textit{MAD} = 3. 

\textbf{Reflection}: As most participants had some prior experience in developing 2D or 3D interactive experiences using either a game engine (e.g., Unity, Unreal) or visual programming (e.g., Max/MSP, Scratch), we asked them to compare their performance and creativity using our system compared with systems they have used. 10 participants rated (average score of 4) that our authoring interface offers an easier way to create cause-effect relationships that can be hard to do with other systems, although one participant (P4) strongly disagreed with this. At the same time, 13 participants agreed that our system was able to help them express their creative ideas well. 11 participants believed that our system could help them be more creative. 
10 participants reported that they would like to use the system for interactive digital storytelling and 11 agreed that the system can support the creation of different types of interactive experiences.


\subsection{In-depth Study Methodology}
The workshop helped us understand the overall learnability of our authoring interface and workflow but it did not demonstrate ConnectVR's performance in real creative practice. To further understand how our system can be applied to a real creative production environment, we conducted an in-depth study with two participants (not from the workshop study) who used our system for one hour per day for a total period of 15 days. Based on this extensive usage, we designed our evaluation to, 1) see if our visual interface and workflow improve their creative process, 2) see if our tool enables the creators to produce new forms of creative experiences that they could not do before, 3) gain insights about how the tool may affect the way they think about creating interactive VR experiences. We recruited two participants from creative backgrounds (1 female and 1 male) to use our system. Like our workshop, we looked for artists, storytellers and creative content producers with a particular interest or prior experience in interactive VR experience design. PA is a media artist and professional UI/UX designer with some experience in immersive and spatial design with Unity. PB is an undergraduate student designer of 3D games in Unity but has no prior experience in VR. Regarding their programming experiences, PA has limited technical knowledge of scripting while PB is more comfortable learning and writing code in Unity. Both participants were paid for approximately 20 hours over the three week period on a \$20/hour basis, resulting in a total of \$400 compensation for each person.  

\subsubsection{Onboarding}
Before starting the study, we provided the participants with instructions to install Unity 2020.3 on their own Windows 10 computers. We also provided technical assistance to the participants in setting up their Oculus Quest 2 VR headsets. Before beginning, the two participants were given a unique participant ID and they provided informed consent. They were asked to fill out a pre-study questionnaire as well as a short pre-study reflection survey to report their existing experience with the process of designing and creating interactive VR experiences (if any) as well as their understanding of how the tools they have used in the past may be related to their creative thinking and practice. The survey is composed of three questions and the participants were asked to provide an answer of 150 – 250 words to each of them. Then we guided the participants through the installation of the ConnectVR in Unity and walked them through the necessary basics of using the ConnectVR system to design and develop an interactive VR experience. This step introduced the participants to the visual interface, the concept of a trigger-action pattern on agent design, as well as some examples of creating multi-modal player interactions based on voice and gestures. 

\subsubsection{Study Task}
The participants were asked to design and deliver an interactive VR experience based on their own creative interests and practice. 
They were given flexibility in prioritizing their creative tasks such as asset preparation, 3D modeling, interaction design, scene building, etc. They were asked to fill out a short survey each time they finished a creative session (typically once a day) regarding their specific creative goals, the difficulty encountered if any, the overall feeling, and other comments that they would like to share. The participants were asked to synchronize the project using Unity’s cloud collaboration tool so that the researchers can examine their daily progress. The researchers collected the survey feedback and project updates, and then followed up with the participants every three days in breif 15-minute Zoom meetings to address any technical issues. 
Similar to the workshop study, the participants received an asset bundle of hundreds of 3D models to help them easily start prototyping and testing but with the option to use their own assets. 

\subsubsection{Offboarding}
After the 15-day period, the participants were invited to finish another short reflection survey on their experience with ConnectVR and its impact on their creativity and creative goals. This was followed by a 20-minute semi-structured Zoom interview to gather insights and feedback on their experience. The questions were designed based on the evaluation guidelines for user interface systems \cite{Olsen2007} and HCI toolkit research \cite{ledo2018}, focusing on expressive leverage, problems not previously solved, targeting population, functionality and general utility feedback. The interviews were screen-recorded after getting consent from the participants.

\subsection{In-depth Study Results}
\begin{figure*}[!t]
\includegraphics[trim={0 2cm 0 0}, clip,width=1\textwidth]{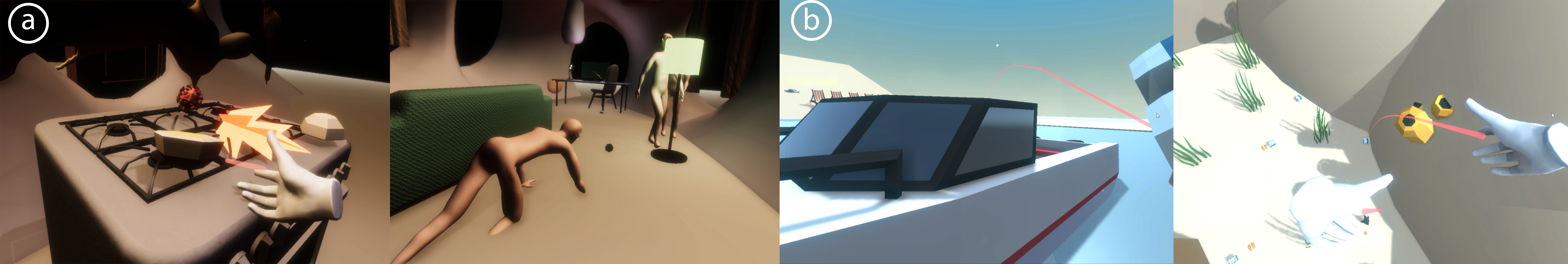}
\abovecaptionskip=1pt
\belowcaptionskip=-12pt
    \caption{VR scenes from the in-depth study results. (a) shows the VR narrative created by PA where the player is placed in a surreal cave-like room filled by daily-life furniture pieces while being surrounded by human-like characters that look the same but behave differently (e.g., talking, crawling, working, sleeping). (b) shows a mini VR game with an environmental theme by PB that the player's mission is to explore an underwater environment of ocean creatures while collecting and bringing back random objects to a boat.}
    \label{fig:pawork}
\end{figure*}

Both participants delivered an interactive VR experience at the end of the 15-day period. We received 11 usage reports from each participant suggesting that they both spent at least 11 sessions with our system during the three weeks and each of these sessions lasted between 1 - 3 hours. PA's narrative focused on a person's mental state of being isolated from external social relationships during COVID period. The narrative was staged in a surreal cave-like room with a number of furniture pieces such as a computer desk, sofa, stove, and lamps. PA modeled and created a number of virtual human characters that looked the same but behaved differently in the room (e.g., talking, crawling, working, sleeping). The player could interact with these clones to reveal hidden aspects of the room such as appearance of mysterious creatures, changes in object appearance, sound effects, etc. PB designed an interactive VR mini-game with an environmental theme staged in an underwater space. The player can move around in a cubic pool to collect various objects such as bottles and boxes, and bring them to a boat to get rewarded. While the player is navigating in the space, they may encounter ocean creatures such as fish and moving grass, and these creatures will notice the player and follow them. \autoref{fig:pawork} shows sample scenes from PA's and PB's work with ConnectVR. Below, we detail the results in the context of our evaluation criteria based on the interview transcripts, usage reports, and reflection write-ups:

\textbf{Process}:
Both participants found that the tool well matched their creative workflows. PA thought that our system fitted well with their creative goal and practice of designing expressive agents with different behaviors. As a UI/UX design professional, in their reflection write-up, PA compared ConnectVR with other graphic design tools they used in their daily work and said: \textit{``after using the tool, I feel the process of creating interactive VR experiences can be simple and straightforward like I do with other kinds of visual designs. It's very convenient to design actions by selecting predefined behaviors and design interactions by connecting each action.''} PB, on the other hand, thought that ConnectVR fitted well with their typical process of creating games that the tool can be easily adopted into their regular workflow to create an interactive experience, replacing the scripting with easy to create causal relationships. They stated in the interview that, ``I was using it in a way as how I would usually go about my scene making process and just try to fit into my workflow instead of using it as an inspirational tool. It turns out that it works pretty well in doing these tasks.''
Both participants thought this tool helped them save a significant amount of time to design interactions. PA thought that the tool helped them overcome difficulties in motions and interactions which used to be \textit{``the most complex part''} for them. But now this tool \textit{``largely reduces the complexity of creating actions''} so that their VR world could become \textit{``more dynamic and interactive.''} PB thought about time efficiency from the point of view that this tool outperformed the scripting part in their workflow. They repeated multiple times over the three weeks that ConnectVR, in comparison to a typical scripting-based workflow, provided a much easier way to create interactions and agent behaviors. They said, \textit{``when I was trying to create, I didn’t really use any scripts or have to open up any scripts to change it, I just use the tool as given…I think it just replaces the scripting part.''} Similarly, when PB was discussing their own practice in the past, they mentioned how they would have done with the traditional way of creating interactive behaviors: \textit{``I think overall it makes a lot of things faster because like I usually have to search things up if I want to make something. If I want to make something follow the player, I might want to search for Unity’s API and have to go from there.''} 

\textbf{Creative Outcomes}: 
PA's previous art practice focused on designing autonomous virtual creatures with animated movements in VR. In the study, PA included many animation clips for their virtual characters and used ConnectVR to author interactive triggers for them. The animations were triggered by different actions by other virtual objects as well as player movement and body-related actions. In this way, they used our system to make their agents more dynamic and responsive to the player and the environment, which had been an unachievable goal until now. In PA's usage reports, PA mentioned that they would like to design \textit{``relations and triggers with a timely order''} but thought that it was a bit \textit{``hard to visualize when different actions happen in a timely order''}. So PA later in their interview, suggested that it would be nice to provide a greater degree of temporal control with a visualized timeline to help design long sequences of actions and know when something will happen.  As PB lacked prior VR experience, they were excited about the fact that they could make an interactive VR experience with very low friction. In PB's total 11 usage reports, they reported 10 times that their creative goal was fully met except for one instance on Day 2 when they encountered a challenge. They had wanted to modify what different VR control buttons do in relation to the player's locomotion behaviors (e.g., fly navigation, teleportation). However, this mapping is managed by Unity's VR input system. After receiving this feedback, the researcher walked them through Unity's VR settings in a follow-up meeting to help solve the problem. 

\textbf{Reflection}: From the reflection survey, we observed that participants were aware of how different creative tools may influence their creative thinking. PA specifically mentioned that \textit{``different creative tools increase my creative thinking in different aspects. This tool encourages me to think more about how different elements in the world can relate to and affect each other. It also helps me to extend my imagination to how VR audiences can participate and influence the virtual environment.''} This parallels PA's creation process during the study as we observed through their usage reports and project updates. PA took an iterative design approach that went back and forth between the agent action design and scene building. They spent a lot of time on making the VR scene and agent actions match with each other for their narrative. In one of their usage reports, they wrote: \textit{``I have been thinking how the room scene is reflecting what's happening inside and transformed over time in a surreal way. I experimented with the claymesh tool to create hidden creatures and design their relation with the original room objects.''}
PB looked at the tool's influence in another way, as an effect brought by time efficiency. They especially emphasized the idea of ``flow'' in their creative process: \textit{“When I create for long periods of time, I fall into a ‘flow state’ where I naturally get progress done without stops. Tools are meant to facilitate that process, instead of stopping the flow.''} In PB’s perspective, the tool must make him able to do fundamental actions faster so that they could \textit{``think of more ambitious ideas''}. PB thought that ConnectVR significantly made a lot of things faster for them, and they could threfore, spend more time thinking about more interactions between different objects for the experience design. They said: \textit{``having a tool that does all this quickly allows me to do all of this in just a few seconds, and lets me test out more interactions because it isn't a time sink to do so.''}

\section{Discussion}
Our results indicate that the authoring interface is generally easy to learn and provides a simple entry point to create agent-based VR narratives for creators and designers without coding experience. Here we discuss the creative opportunities that resulted from the design of our system through three main questions: 1) Was ConnectVR compatible with the workflow of non-technical creators? 2) Did ConnectVR offer meaningful creative opportunities? 3) What are the limitations of the trigger-action method for creating VR narratives? 

\subsection{Was ConnectVR compatible with the workflow of non-technical creators?}
ConnectVR was designed specifically based on the idea that an agent-based simulation approach would require little domain knowledge so that it would be easy to learn and use for non-technical creators. From the in-depth study, we observed that both PA and PB were able to fulfill their creative objectives. The projects they worked on were consistent with their own creative practices such that PA could expand on their artistic pursuit of creating animated virtual agents in their VR narrative composition and PB could implement their game design ideas as a VR experience. Both PA and PB described ConnectVR's trigger-action pattern as an ``easier'' and ``faster'' method for creating interactive VR experiences compared to their prior experience with scripting and other visual programming tools. This was especially true for PA who mentioned in their reflection survey that they did not have many options for creating VR interactions, by saying: \textit{``to achieve a complex interaction [it] will require me to write code in C\# and this could limit my thinking and creation of interactive VR experience.''}. For PA, ConnectVR's simple and fast no-code approach was very useful and it allowed them to expand their art practice into a new domain without anyone's assistance. 
This also resonated with PB's thoughts on the benefit of a faster workflow in allowing them to spend more time on designing the interactivity of an experience. The results suggest that the intuitive, no-code nature of ConnectVR is as important as the features it enables, i.e., design of cause-effect links. This is because creators without technical expertise may want to spend more time on the creative aspects of VR development and less time and effort on the technical aspects. Complex interfaces and software can present a barrier between the creator and their vision, requiring them to spend considerable time learning and navigating the tool's interface to access the functionality they need. This can stifle creativity or discourage non-technical users from creating with the tool or even exploring a new domain such as VR. By removing the complexity of scripting or the tooling interface, creators can experience a more positive and efficient creative workflow. 
\vspace*{-2pt}
\subsection{Did ConnectVR offer meaningful creative opportunities?}
A common concern about designing a tool with greater accessibility is that it may have a lower expressiveness. However, the results of our studies indicate that ConnectVR's accessibility did not constrain its ability to support the users in creating different types of experiences. From the workshop study, the participants created experiences that showcase diverse narrative themes such as a treasure hunt (P6), a zombie breakout (P12), lives of church goers (P15), and workers in protest (P3). The narratives are presented in different formats such as an art experience (P11, P1), a generative system (P9) and a mini-game (P14). P11 created a fantasy land occupied by unusually large sized flowers and mushrooms where a player's movement triggered different behaviors on these objects such as dancing, singing, and changing scene lighting (Figure~\ref{fig:results}). P14 created a game-like story where a T-Rex chases a group of people and the player's goal, hinted at by a virtual character, is to left click the mouse to shoot pineapples at the T-Rex to stop it in its path. P9 designed a generative mechanism using the provided \textit{Generate} command to procedurally spawn interactable agents upon each action trigger. In P9's story, a player encounters a COVID-like virus in the forest. This virus object duplicates itself and spreads upon being approached by animal or human characters. This mechanism, simulating a possible real world scenario, made it hard for P9 to predict how far and how much the virus would spread over the story world, which they thought was interesting. These examples of interactive relationships were unexpected and we believe support the idea that our system can allow for flexibility in approach to support creativity.

During the studies, we noticed frequent use of actions related to player movement or speech, leading to a style of narratives composed largely of the player exploring the virtual environment and engaging in conversations with virtual characters. Virtual agents that appeared in 13 stories were designed similarly to gauge a player's movement or identify attention (when the player looks at them) to automatically start a conversation with the player or trigger other story elements. We speculate that participants preferred using natural and familiar ways of interaction as triggers for the narrative events in their introductory stories and considered such natural interactions well suited for creating engaging experiences. This was not unlike the popular genre of JRPG games that follows a similar conversational narrative pattern, but without animated visuals. Based on these observations and the creative projects, we believe that by integrating a wide range of player input methods into player actions, ConnectVR can help creators tell new types of stories in VR.

\subsection{What are the limitations of the trigger-action approach?}
While participants were positive on the ease of using our trigger-action method to create cause-effect chains, there are some limitations we observed during our studies. In ConnectVR, getting accurate temporal control on agent behaviors can be difficult. 
While our system offers controls for time-based interactions in VR, the time it takes for certain actions to play out can be difficult to synchronize with other parallel actions. This is because actions may consist of multiple commands, some of which may take unpredictable amounts of time to execute (such as generating speech with a language model or moving to a specific location). 
P4 and PA reported this limitation, that it was hard to synchronize certain time-based actions with other actions. They suggested adding timeline features to visualize the temporal information to help with the synchronization process. While ConnectVR's agent-based simulation approach has advantages for generating procedural behaviors on the fly, in its current form it is less well-suited to precise planning and scripting of events. 

Another limitation of ConnectVR comes from the unpredictable nature of agent actions. While our system makes it easy to create chains of actions, our participants noted that it can be challenging to design and think of plausible effects over long periods of time. However, that may not be desirable either as it's not meaningful to a player's experience when an action happens in another part of the scene that they cannot see or be part of. 
The longest chained relationship we observed in our studies was in P10's story depicting a gun fight between two cowboys using seven chained actions. In their scenario,  the player is confronted by a gun fight between two cowboys and is given two options, either to stand by the one who is the husband of a lady or stand by his rival. Choosing the rival triggers a sequence of actions (the husband dies, the lady screams, horses escape and run into wooden crates that explode) and consequences that end with the burning down of the village. Most of the chained relationships we observed in the workshop study were not very long (\textit{M} = 3 actions in each chained relationship). Our intuition is that the design of a long sequence of cause-effect relationships may require careful planning of the story and reasoning of interactions between different agent behaviors, which can take a lot more time than the participants had during the workshop, and significantly more effort at the conceptual level, much like making a movie with such complex sequences of events.

\section{Conclusion}
In this paper, we presented ConnectVR, a trigger-action authoring interface to enable non-technical creators to design VR narrative experiences driven by complex cause-effect relationships between a player’s actions and corresponding agent behaviors. We presented details of our system design and features that support the 3D and VR narrative authoring process. A workshop study with 15 participants and a three-week in-depth study with two creators evaluated the usability, creativity support, and ease of building the trigger-action graph using the visual interface. Our results show that participants were able to build their own narrative experiences using ConnectVR and they found it easy and intuitive to use for creative storytelling. The use of a trigger-action graph to define actions and cause-effect relationships was quite successful with evidence indicating that the large variety of player-driven actions can support new types of expressive VR storytelling.

\bibliographystyle{abbrv-doi}

\bibliography{references}
\end{document}